\documentclass[ALICE,manyauthors]{cernphprep}
\usepackage[comma,square,numbers,sort&compress]{natbib}
\usepackage{hyperref}
\usepackage{lineno}
\usepackage{inputenc}
\usepackage{fontenc}
\usepackage{hyperref}
\usepackage{graphicx}  
\usepackage{dcolumn}   
\usepackage{bm}        
\usepackage{amssymb}   
\usepackage{amsfonts}
\usepackage{graphics}
\usepackage{grffile}   
\usepackage{epsfig}
\usepackage{units}
\usepackage[usenames]{color}
\usepackage[normalem]{ulem} 
\usepackage{lineno}
\usepackage{color}
\usepackage{subfigure}
\usepackage[T1]{fontenc} 

\newcommand{ \be }{\begin{linenomath*}\begin{eqnarray}}
\newcommand{ \ee }{\end{eqnarray}\end{linenomath*}}
\newcommand{ \la }{\langle}

\newcommand{ \ra }{\rangle}

\def \mean#1 {{\la #1 \ra}}

\newcommand{ \PbPb}{Pb -- Pb}

\definecolor{dgreen}{cmyk}{1.,0.,1.,0.2}        
\definecolor{orange}{cmyk}{0.,0.353,1.,0.}    

\begin{document}%

\begin{titlepage}
\PHyear{2021}
\PHnumber{261}      
\PHdate{09 December}  
%

\title{Neutral to charged kaon yield fluctuations in \PbPb\ collisions at $\sqrt{s_{\rm NN}}$~=~2.76~TeV}
\ShortTitle{Neutral to charged kaon yield fluctuations}   

\Collaboration{ALICE Collaboration\thanks{See Appendix~\ref{app:collab} for the list of collaboration members}}
\ShortAuthor{ALICE Collaboration} 

\begin{abstract}
We present the  first measurement of event-by-event fluctuations in the kaon sector in \PbPb\ collisions at $\sqrt {s_{\rm NN}}=$ 2.76 TeV with the  ALICE detector at the LHC. The robust fluctuation correlator $\nu_{\rm dyn}$ is used to evaluate the magnitude of fluctuations of the relative yields of neutral and charged kaons, as well as the relative yields of charged kaons, as a function of collision centrality and selected kinematic ranges. While the correlator $\nu_{\rm dyn}[\rm K^+,\rm K^-]$ exhibits
a  scaling  approximately in inverse  proportion of the charged particle  multiplicity, $\nu_{\rm dyn}[\rm K_S^0,\rm K^{\pm}]$ features a significant deviation from such scaling. Within uncertainties, the value of  $\nu_{\rm dyn}[\rm K_S^0,\rm K^{\pm}]$ is independent of the selected transverse momentum interval, while it exhibits a pseudorapidity dependence. The results are compared with HIJING, AMPT and EPOS--LHC predictions, and are further discussed in the context of the possible production of disoriented chiral condensates in central \PbPb\  collisions.  
\end{abstract}
\end{titlepage}
\setcounter{page}{2}

The primary intent of high-energy heavy-ion collisions at the Relativistic Heavy Ion Collider (RHIC) and at the Large Hadron Collider (LHC) is the production and study of the state of matter  in which  quarks and gluons are deconfined. The matter formed in these collisions is characterised as a low-viscosity fluid, which undergoes a transition to the hadronic phase, after it expands and cools down~\cite{doi:10.1146/annurev-nucl-101917-020852}. The transition  from the hadronic to the QGP phase involves a  partial restoration of chiral symmetry and color deconfinement. Deconfinement of quarks and gluonic degrees of freedom, and the 
production of QGP  were brought to light by measurements of jet quenching~\cite{Adler:2002tq,Adare:2008ae,PhysRevLett.97.162301,PhysRevC.77.011901,PhysRevC.75.034901}, quarkonium state suppression, and measurements of anisotropic flow~\cite{Adler:2002pu,AlversRoland,Abelev:2007qg,PhysRevLett.105.252302,Aamodt:2010pa,Adam:2017ucq}. Anomalous fluctuations of conserved charges were 
predicted to arise in the vicinity of the phase boundary and potential signals of the production of a deconfined phase~\cite{Stephanov:2004wx,Bazavov:2011nk,Friman:2011pf,Borsanyi:2018grb,Redlich:2007ud}. 

Several studies of particle yield fluctuations have already been reported~\cite{Pruneau:2003ix, Pruneau2004, PhysRevC.64.041901, PhysRevC.64.051902} but their interpretation is largely a matter of debate. More recently, the ALICE collaboration has also investigated fluctuations of net charge, net protons, as well as fluctuations of the relative yields of pions, kaons, and protons~\cite{Abelev:2012pv,Acharya:2019izy,Acharya:2017cpf}. Measurements of such fluctuations are of interest, in particular, as they are nominally  sensitive to QGP susceptibilities~\cite{BRAUNMUNZINGER2016805}, the proximity of the hadron gas-QGP phase boundary, as well as, when considered at lower beam energy, the existence of a critical point~\cite{PhysRevC.98.024910,PhysRevLett.81.4816} in the phase diagram of nuclear matter. Additionally, measurements of fluctuations are also of interest to probe the existence (or proximity) of the chiral phase transition, which should manifest itself by the production of anomalous fluctuations~\cite{Redlich:2007ud}. A specific manifestation of this transition from the chiral symmetric phase (high temperature) to a broken phase (low temperature) involves the production of disoriented chiral condensates (DCCs)~\cite{RAJAGOPAL1993395}, a region in isospin space where the chiral order parameter is misaligned from its vacuum orientation. Theoretical studies of the production and decay of DCCs are typically formulated in the context of
the SU(2) symmetry. It is predicted that the production and decay of DCCs shall manifest through enhanced fluctuations of neutral and charged pion multiplicities~\cite{PhysRevD.46.246,HiroOka:1999xk}. The past searches in this sector have yielded no evidence for DCC production~\cite{PhysRevD.61.032003,0954-3899-23-12-009,Adamczyk:2014epa}. However, the production of distinct DCC domains might result in ``isospin fluctuations'' in the kaon sector, i.e., enhanced fluctuations of the relative yields  of neutral and charged kaons which can be measured by means of the $\nu_{\rm dyn}[\rm K^0, \rm K^c]$~\cite{PhysRevLett.86.4251,Gavin:2001uk,2003NuPhA.715..657A,Gavin:2002ci,Pruneau:2002yf}. Specifically, it was predicted that the production of DCC domains in  A--A collisions might lead to an anomalous scaling of the net charge correlator $\nu_{\rm dyn}$, defined in the following, as a function of charged particle multiplicity~\cite{PhysRevC.101.054904}. A search of kaon isospin fluctuations at LHC energies is thus of significant interest. Given that kaons from DCC are expected to be produced at modest transverse momentum~\cite{GAVIN1995163}, the present study is restricted to measurements of $\rm K^{\pm}$ and $\rm K_{S}^{0}$ at the lowest possible transverse momenta ($p_{\rm T}$) aiming at checking whether data support some basic expectations from the DCC production.

A measurement of dynamical neutral-to-charged kaon fluctuations, with $\nu_{\rm dyn}$, is also of interest  in the broader context of two-particle correlations induced by the hadronization of the QGP, via e.g., quark coalescence, and the transport of produced hadrons, as well as the possibility that high mass resonances lead to the production of pairs of  kaons. Examples of such resonance decays that  contribute to the $\nu_{\rm dyn}$ correlator include $\rm \phi(1020)\rightarrow \rm K^{+} +\rm K^{-}$, $\rm \phi_3(1850)\rightarrow \rm K + \overline{\rm K}^*$, $f_2(2300)\rightarrow \rm \phi+ \rm \phi$, as well as several $\rm D$-mesons states. 
As they decay in-flight, these high-mass states would  induce pair correlations of charged and neutral kaons. The relative abundance of such states might be larger in central collisions because of higher initial temperature and density conditions. The relative yield of neutral and  charged kaons, and their fluctuations, might then exhibit a centrality dependence  as a result of feed-down contributions from such states. Additionally, given that the relative yield of neutral kaons and charged kaons in general is  determined by the relative yields of strange, up, and down quarks (and their anti-particles)  before hadronization, and given that the production of strangeness is both energy and collision centrality dependent, one might anticipate a change in the size of the fluctuations from peripheral to central heavy-ion collisions. Note, however, that the presence of kaons resulting from feed-down contributions, which, for central \PbPb\ collisions at LHC energy, amount to about 50 percent at the lowest momenta considered~\cite{Andronic:2017pug}, reduces but does not eliminate the sensitivity of the method to the presence of strange DCCs or other processes inducing  variations of the direct kaon production~\cite{PhysRevC.101.054904}.

In this letter, we report  measurements of event-by-event fluctuations  of  inclusive multiplicities of charged and neutral kaons based on the robust statistical observable, $\nu_{\rm dyn}$, defined as 
\begin{equation}\label{eq:nudyn}
\nu_{\rm dyn}=R_{\rm cc}+R_{\rm 00}-2R_{\rm c0},
\end{equation}
where the indices $\rm c$ and $0$ stand for charged  and neutral kaons, respectively.
The correlators $R_{xy}$ are normalized factorial cumulants  calculated  according to 
\begin{equation}
R_{xy}=\frac{\left<N_{x}(N_{y}-\delta_{xy})\right>}{\left<N_{x}\right>\left<N_{y}\right>}-1,
\end{equation}
Here $\delta_{xy} =$ 1 for x = y and 0 for $x \ne y$. The correlators $R_{xy}$ and $\nu_{\rm dyn}$  vanish
in the absence of pair correlations, i.e., for Poisson fluctuations, but deviate from zero in the presence of particle correlations. Their magnitudes are
expected to approximately scale in inverse proportion of the charged particle  multiplicity, $N_{\rm ch}$, produced in heavy-ion collisions and shall be   insensitive to detection inefficiencies, and only weakly dependent on the acceptance of the measurement~\cite{Pruneau:2002yf}.

In order to reduce the challenge of measuring neutral kaon multiplicities on an event-by-event basis, the neutral kaon measurement is restricted to $\rm K_{\rm S}^{0}$ by means of their decay into a pair of $\rm \pi^+$, $\rm \pi^-$ (69.2$\%$ branching ratio~\cite{Zyla2020}) with a displaced vertex. The wide acceptance and high detection efficiency of charged pions enables event-by-event reconstruction of $\rm K_{\rm S}^{0}$ with high efficiency
and small combinatorial background. It is thus possible to measure their multiplicity event-by-event and compute the first, $\langle N_{\rm K_{\rm S}^{0}}\rangle$, as well as second, $\langle N_{\rm K_{\rm S}^{0}}( N_{\rm K_{\rm S}^{0}} -1) \rangle$ factorial moments.   These  constitute  estimators to  moments of neutral kaon (and anti-kaon) yields. Indeed, given neutral kaons
have a  50\% probability of being a $\rm K_{\rm S}^{0}$, with a binomial probability distribution, a measurement of the ratio $\langle N_{\rm K_{\rm S}^{0}} (N_{\rm K_{\rm S}^{0}}-1)\rangle/\langle N_{\rm K_{\rm S}^{0}}\rangle^2$ is thus strictly equivalent to $\langle N_{\rm K^{0}} (N_{\rm K^{0}}-1)\rangle/\langle N_{\rm K^{0}}\rangle^2$. Likewise, $\langle N_{\rm K_{\rm S}^{0}} N_{\rm K^{c}})\rangle/\langle N_{\rm K_{\rm S}^{0}}\rangle \langle N_{\rm K^{c}} \rangle$ is equivalent to 
$\langle N_{\rm K^{0}} N_{\rm K^{c}})\rangle/\langle N_{\rm K^{0}}\rangle \langle N_{\rm K^{c}} \rangle$. A measurement of $\nu_{\rm dyn}[N_{\rm K_{\rm S}^{0}},N_{\rm K^{c}}]$ thus provides a proper and unbiased proxy to that of $\nu_{\rm dyn}[N_{\rm K^{0}},N_{\rm K^{c}}]$ even without a measurement of $\rm K_{L}^{0}$.

The results presented in this letter are based on  $1.3\times10^{7}$ minimum bias (MB) \PbPb\ collisions  at center-of-mass energy per nucleon pair, $\sqrt{s_{\rm NN}}$ = 2.76 TeV collected during the 2010 LHC  heavy-ion run with the ALICE detector.
The reported correlation functions are measured for charged particles reconstructed within the Inner Tracking System (ITS)~\cite{Nouais:2001cd} 
and  the Time Projection Chamber (TPC)~\cite{Baechler:2004ge}. 
The TPC  consists of a 5 m long  gas volume contained 
in a cylindrical electric field cage oriented along the beam axis, which is  housed within a large solenoidal magnet designed and operated 
to produce a uniform longitudinal magnetic field of 0.5 T. Signals from charged particles produced in the TPC gas are readout at both end caps. The ITS is comprised of three subsystems, each consisting of two cylindrical layers of
silicon detectors designed to match the acceptance of the TPC and provide high position resolution.
Together, the TPC and ITS  provide  charged particle track 
reconstruction and momentum determination with full coverage in azimuth over the pseudorapidity range $|\eta|<0.8$  and with good reconstruction efficiency for charged particles with $p_{\rm T}>0.2$ GeV$/c$. Two forward scintillator
systems, known as  V$0$A  and V$0$C, covering the pseudorapidity ranges $2.8 < \eta < 5.1$ and  $-3.7 < \eta < -1.7$, respectively, are additionally used for triggering and event classification purposes.  Detailed descriptions of the ALICE detector components and their respective performances are given in Refs.~\cite{Aamodt:2008zz,Abelev:2014ffa}.
The MB interaction trigger required at least two out of the following three conditions: i) two pixel chips hit in the outer layer of the silicon pixel detectors ii) a signal in V0A iii) a signal in V0C. The hit multiplicity in the V0 detectors is additionally used to estimate the collision centrality  reported in seven  classes corresponding to 0--5\% (most central), 5--10\%, 10--15\%, 15--20\%, 20--40\%, 40--60\%, and  60--80\% (most peripheral)  of the hadronic \PbPb\ cross section~\cite{Abelev:2013qoq}.  The approximate position along the beam line of the primary vertex ($z_{\rm vtx}$) of each collision is first determined based on hits recorded in the two inner layers of the ITS detector.  Reconstructed charged particle tracks in the ITS and TPC are finally propagated to the primary vertex to achieve optimal position resolution. In the context of this analysis, the vertex  is required to be in the range $|z_{\rm vtx}| \le 10$ cm of the nominal interaction point 
 in order to ensure a uniform detector acceptance and  minimize variations of the efficiency across the fiducial volume of the experiment. In addition, the standard track quality selections were used to ensure
that only well-reconstructed tracks were taken in the analysis. All selected tracks of each event are processed to identify pions and kaons with the techniques described below.  Event-by-event combinations of two oppositely charged pions are formed  to reconstruct topological $\rm K^0_{\rm S}$ candidates. Standard ALICE 
topological criteria, also detailed below, are then used to identify and select $\rm K^0_{\rm S}$ candidates.
 Charged and neutral kaons are counted, event-by-event, to  calculate the moments 
$\langle N_{c}\rangle$,  $\langle N_{0}\rangle$,  $\langle N_{c}(N_{c} -1) \rangle$,
 $\langle N_{0}(N_{0} -1) \rangle$ and $\langle N_{c} N_{0} \rangle$,  in each collision centrality class. The corrections for particle losses are obtained event-by-event by dividing single and pair yields by the detection efficiency and products of efficiencies, respectively. Transverse momentum and pseudorapidity dependent efficiencies were evaluated from GEANT simulations (discussed below) of the particle detection performance with the HIJING model. The 
moments are finally combined to calculate $\nu_{\rm dyn}$ values in each class according to Eq.~(\ref{eq:nudyn}).

Charged particle identification (PID) is performed in the pseudorapidity range $|\eta| <  0.5$ using the $n\sigma$ method based on their  energy loss (${\rm d}E/{\rm d}x$) and their time of flight, measured  in the TPC  and TOF detectors, respectively~\cite{Aamodt:2008zz,Abelev:2014ffa}. A selection resulting in a PID with a high purity is crucial in order to minimize biases in measurements of  $\nu_{\rm dyn}$. 
Kaons are selected from TPC ${\rm d}E/{\rm d}x$ with $|n\sigma| < 2$ in the  ranges 0.2 $< p < $ 0.39 GeV$/c$ and 0.47 $< p < $ 0.5 GeV$/c$,  and $-0.5< n\sigma < 2$ in the range 0.39 $< p < $ 0.47 GeV$/c$ to reduce contamination from electrons. Both TPC and TOF signals, with  $|n\sigma| < 2$,  are used in the range 0.5 $< p < $ 1.5 GeV$/c$.  Furthermore,   kaon tracks are selected based on their distance of closest approach (DCA) to the collision primary vertex in order to select primary tracks and suppress contamination from secondary particles and processes. Only tracks with DCAs smaller than 3.2 cm and 2.4 cm along and transverse to the beam direction, respectively, are included in the analysis. These selection cuts lead to charged kaon contamination ranging from 1.0\%  (TPC) to 1.4\% (TPC+TOF) in peripheral collisions, and 2.7\% (TPC) to 4.4\% (TPC+TOF) in 5\% most central collisions. 
 
\begin{figure}
\centering
\includegraphics[width=0.5\linewidth,trim={2mm 2mm 2mm 2mm},clip] {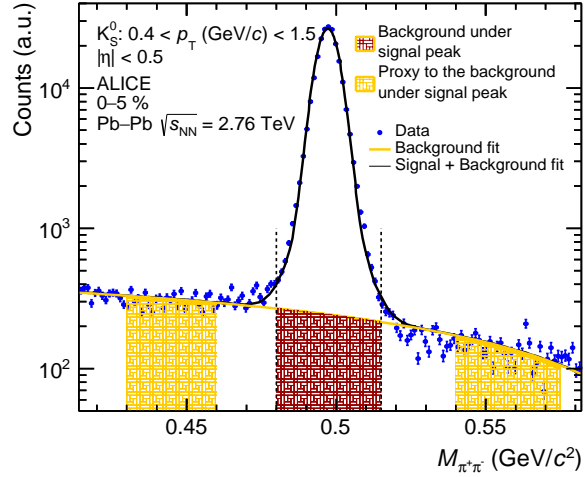}
\caption{Invariant mass distribution of $\pi^++\pi^-$ pairs measured in central (0--5$\%$) \PbPb\ collisions. The yellow and black solid lines show a second order polynomial fit of the combinatorial background and a Gaussian+second order  polynomial fit to the invariant mass spectrum, respectively. The vertical dash lines  delineate the mass range used for the determination of neutral-kaon yields. Given the red-brown area, which corresponds to combinatorial background, cannot be properly assessed event by event, the average of the yellow areas is used  as proxy, event-by-event, to estimate the true combinatorial yield represented by the red-brown area. }
\label{fig:PID:TPC:mass}
\end{figure}

Neutral kaons, $\rm K_{\rm S}^{0}$, are reconstructed and selected within 0.4 $< p_{\rm T} < $ 1.5  GeV$/c$  and  $|\eta|<$ 0.5 based on  their weak decay
$\rm K_{\rm S}^{0}\rightarrow \pi^{+}+\rm \pi^{-}$ topology
and an 
invariant mass selection criterion, 0.480 $< M_{\rm \pi^{+} \rm \pi^{-}}<$ 0.515 GeV$/c^2$ with their decay-product tracks within the acceptance window $|\eta|<$ 0.8.  Standard 
ALICE topological cuts~\cite{PhysRevLett.111.222301} are used towards the selection of  $\rm K_{\rm S}^{0}$ candidates formed from  $\rm \pi^{+}$ and $\rm \pi^{-}$ tracks identified with the TPC and TOF detectors
with $p_{\rm T} > $ 0.2 GeV$/c$. The maximum DCA of neutral kaons is set to 0.1 cm in all directions.  The required 
$n\sigma$ values for the pions were $|n\sigma| < 2$ in both TPC and TOF detectors for 0.2 $< p < $ 1.5  GeV$/c$. These 
selection criteria yield a combinatorial $\rm \pi^++\rm \pi^-$ pair 
contamination ranging from 1.3\% in peripheral collisions to 4.0\% in 5\% most central collisions,  shown as a
red-brown area in Fig.~\ref{fig:PID:TPC:mass}.

Event-by-event fluctuations of the combinatorial background artificially increase the factorial moment $\langle {N_0(N_0-1)}\rangle$
and may bias $\langle N_0 N_c\rangle$. A correction for such contamination is accomplished by additionally measuring correlators 
involving  moments of the number of background pion pairs, $N_{\rm b}$, in side band mass ranges
$0.430 < M < 0.460$ GeV$/c^2$ and $0.540 < M < 0.575$ GeV$/c^2$ used as proxies of the number of background pairs in the nominal mass range $0.480 < M < 0.515$ GeV$/c^2$, shown in Fig.~\ref{fig:PID:TPC:mass}, as yellow and red-brown areas. 
The background suppressed correlators $R_{\rm 00}$ and $R_{\rm c0}$ are thus calculated according to 
\be\label{eq:correctionContamination1}
R_{\rm 00}&=& (1-f_{ab})^{-2} \left[ R_{\rm aa} - 2f_{ab} R_{\rm ab} + f_{ab}^{2} R_{\rm bb}\right], \\
\label{eq:correctionContamination2}
R_{\rm c0}&=& (1-f_{ab})^{-1}\left[R_{\rm ac} -f_{ab} R_{\rm bc}\right],
\ee
where the labels $\rm a$, $\rm b$, and $\rm c$ represent pairs in the nominal mass range,  
pairs observed in either side bands, and pairs of charged kaons, respectively. The fraction $f_{ab} = \la N_{\rm b}\ra/\la N_{\rm a}\ra$ is determined for each
collision centrality bin as the average number of background pairs in the range $0.480 < M < 0.515$ GeV$/c^2$
estimated from a polynomial fit to the background, as illustrated in Fig.~\ref{fig:PID:TPC:mass}. Corrections for combinatorial contamination based on Eqs.~(\ref{eq:correctionContamination1},\ref{eq:correctionContamination2}) range from 4.7\% in peripheral to 2.5\% in central collisions.
Additionally, given the number of charged and neutral kaons grows monotonically with collision centrality, values of $\nu_{\rm dyn}[\rm K_{\rm S}^{0},\rm K^{\pm}]$ are  corrected for finite centrality bin widths. The bin width correction is calculated by considering the weighted average of $\nu_{dyn}$ evaluated in 1\% intervals of collision centrality across the reported bin widths~\cite{Luo_2013}. These corrections range from 3.9\% in peripheral to 2.1\% in central collisions.

Statistical uncertainties are evaluated with the event subsampling method using 10 subsamples~\cite{Acharya:2017cpf,Abdelwahab:2014yha}. 
The systematic uncertainties include contributions from secondary particles, as well as from the $p_{\rm T}$ dependence of the tracking efficiency which is not perfectly canceled in the determination of the normalized cumulants $R_{xy}$. The event and track selection criteria were varied, and a statistical test~\cite{Barlow:2002yb} was used to identify significant sources of systematic uncertainties. The largest sources of systematic uncertainties include: (i) the effect of varying the minimum or maximum decay length ($<4$\%), (ii) variations of the K$^{\pm}$ purity when changing the sigma selection criteria ($<4$\%), (iii) variations during the data taking period of the $\rm K^{\pm}$  and $\rm K_{\rm S}^{0}$ yields ($<3$\%),  and (iv) variations of $R_{00}$ when correcting for combinatorial background using different fiducial invariant mass  ranges and side bands ($<2$\%). Additional uncertainties arise when  varying the range of accepted primary vertices ($<4$\%). Adding all sources in quadrature, 
 systematic uncertainties are estimated to be smaller than 13\%, independently of collision centrality.

\begin{figure}
\centering 
\includegraphics[width=0.48\linewidth,trim={2mm 2mm 2mm 2mm},clip] {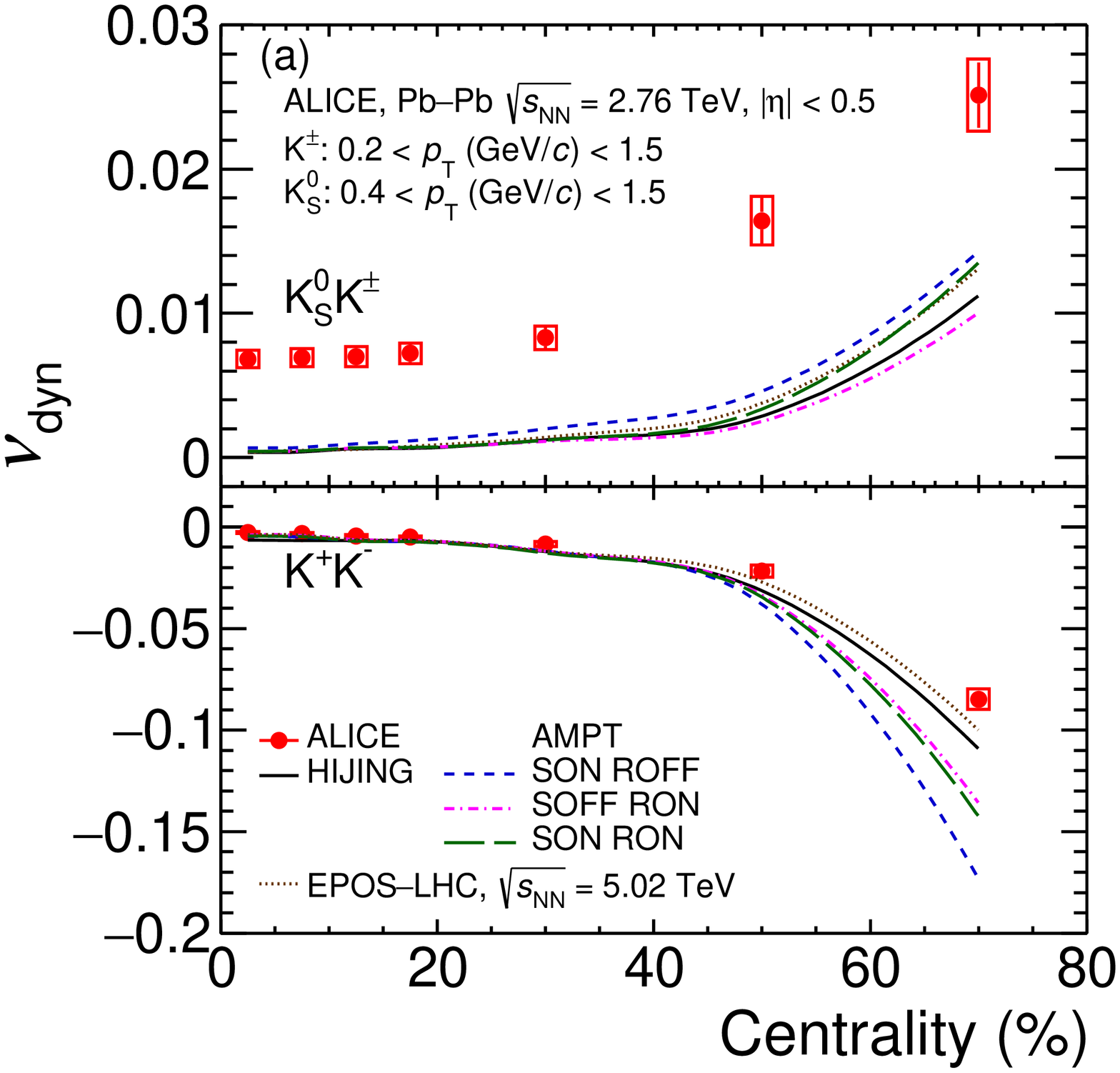}
\includegraphics[width=0.48\linewidth,trim={3mm 2mm 2mm 2mm},clip] {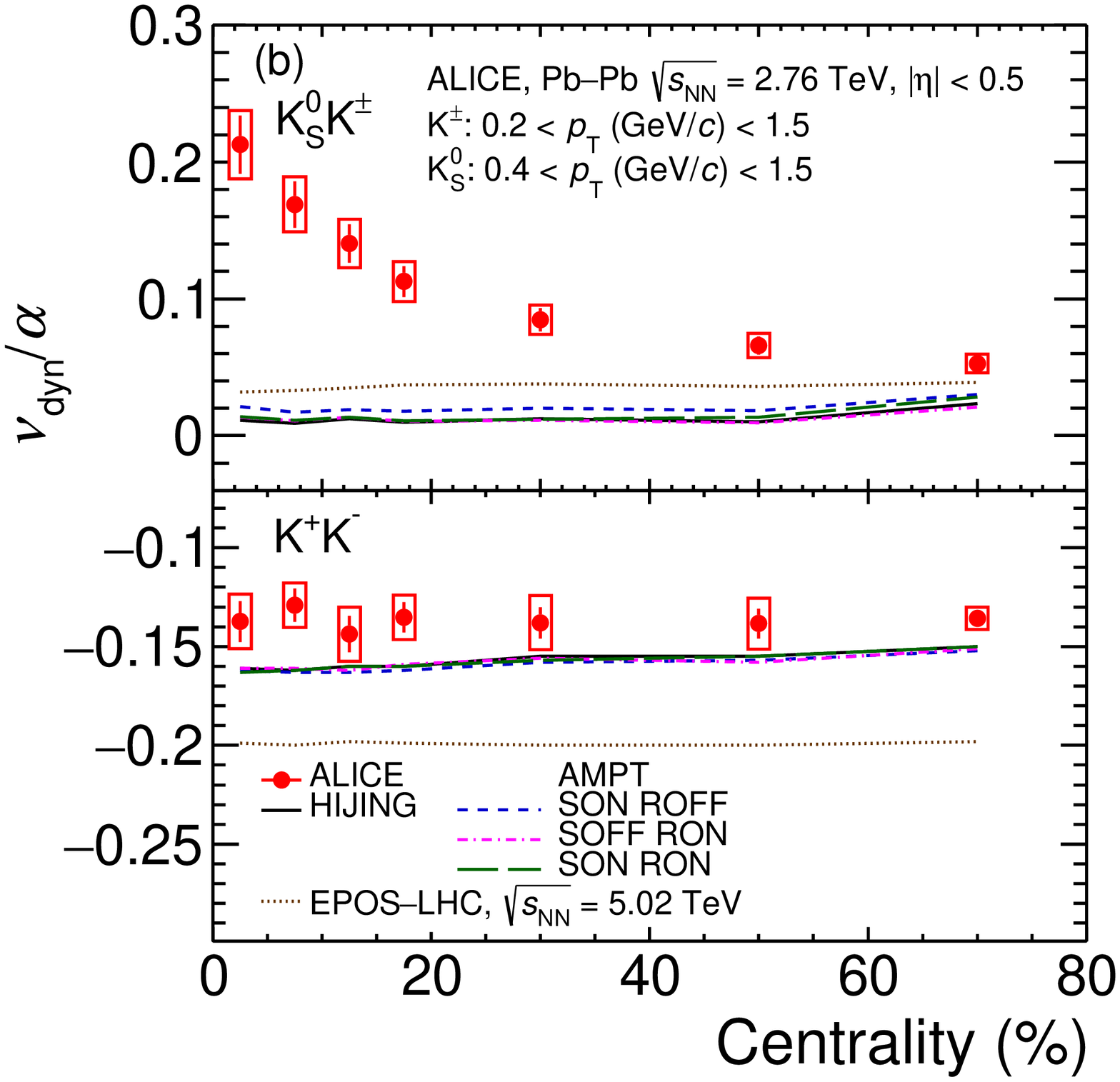}
\caption{
(a) Measured values of $\nu_{\rm dyn}[\rm K_{\rm S}^{0},\rm K^{\pm}]$ (top) and $\nu_{\rm dyn}[\rm K^{+},\rm K^{-}]$ (bottom) compared with HIJING and AMPT model calculations of these observables at generator level. (b) Values of $\nu_{\rm dyn}[\rm K_{\rm S}^{0},\rm K^{\pm}]$ (top) and $\nu_{\rm dyn}[\rm K^{+},\rm K^{-}]$ (bottom) scaled by $\alpha \equiv (\la \rm K_{\rm S}^{0}\ra^{-1} + \la\rm K^{\pm}\ra^{-1})$. Statistical and systematic uncertainties are represented as vertical bar and boxes, respectively.
}
\label{fig:NuDynVsCent}
\end{figure}

Heavy-ion collisions simulated with the HIJING v2.0~\cite{PhysRevD.44.3501}  and AMPT v2.21~\cite{Lin:2004en} Monte Carlo (MC) event generators, including the propagation of the produced particles through the detectors using GEANT3 ~\cite{Brun:1994aa}, are used to validate the correction method. To that end, the reconstructed values of $\nu_{\rm dyn}$ obtained from full MC simulations are processed as data, then the fully corrected $\nu_{\rm dyn}$ values are compared to those obtained at generator level, i.e., from simulations without detector effects. The agreement between the reconstructed  and generator level values of $\nu_{\rm dyn}$  are found to be within  1\%. Generator level MC simulations performed with the AMPT and EPOS--LHC~\cite{PhysRevC.92.034906} models are additionally used to obtain basic theoretical expectations for the magnitude of charged to neutral kaon yield fluctuations. AMPT events produced with the options of string melting on (SON) and rescattering off (ROFF), string melting off (SOFF) and rescattering on (RON), and SON and RON are considered.
 Furthermore, EPOS--LHC events are analyzed at generator level for \PbPb\ collisions at $\sqrt{s_{\rm NN}}$ = 5.02 TeV. All  sets of simulated data are analyzed with selection parameters and conditions identical to those used in the analysis of the experimental data. Additionally, the robustness of the analysis was tested by performing a closure test. To that end, values of  $\nu_{\rm dyn}[\rm K_{\rm S}^{0},\rm K^{\pm}]$ obtained with HIJING simulated events at both the detector and generator levels are compared, and verified that values of $\nu_{\rm dyn}$ obtained with full simulations of the detector performance and data reconstruction (detector level) are in excellent  agreement with those obtained with generator level data sets.

The top panel of Fig.~\ref{fig:NuDynVsCent} (a) presents $\nu_{\rm dyn}[\rm K_{\rm S}^{0},\rm K^{\pm}]$ (red solid circles) as  a function of the  Pb--Pb collision centrality. The largest $\nu_{\rm dyn}$ is observed in the most peripheral collisions and monotonically decreases for more central collisions. Such a behavior  is qualitatively  well described
by HIJING, AMPT and EPOS--LHC calculations, but these models
underestimate the magnitude of $\nu_{\rm dyn}[\rm K_{\rm S}^{0},\rm K^{\pm}]$ by an approximate factor of two in peripheral collisions and by an order of magnitude in most central collisions. The magnitude of  $\nu_{\rm dyn}[\rm K_{\rm S}^{0},\rm K^{\pm}]$ is expected to approximately scale in inverse proportion to the number of sources of correlated particles, $N_{\rm s}$,  for a collision system involving independent nucleon--nucleon collisions and no scattering of produced particles. This scaling is explored in the top panel of Fig.~\ref{fig:NuDynVsCent}(b) which displays the centrality dependence of $\nu_{\rm dyn}$ scaled by  the factor $\alpha \equiv (\la \rm K_{\rm S}^{0}\ra^{-1} + \la\rm K^{\pm}\ra^{-1})$. This factor is known to be approximately proportional to $N_{\rm s}$~\cite{Gorenstein:2011hr}. Scaled values predicted by the models are found to be essentially  invariant with collision centrality. HIJING, in particular, exhibits   nearly constant values of $\nu_{\rm dyn}[\rm K_{\rm S}^{0},\rm K^{\pm}]/\alpha$ as a function of collision centrality, whereas AMPT calculations with SON and ROFF options display a very modest collision centrality dependence, hardly visible in Fig.~2(b).  The measured data, by contrast, feature a strong variation with decreasing centrality. In particular, $\nu_{\rm dyn}[\rm K_{\rm S}^{0},\rm K^{\pm}] /\alpha$ rises from $\approx 0.053\pm 0.005 (\rm stat) \pm 0.007 (\rm sys)$ in the 60--80\% collision centrality range to $0.213\pm 0.021 (\rm stat)\pm 0.025 (\rm sys)$ in 5\% most central collisions and thus one then concludes the expected   $1/N_{\rm s}$ scaling  of $\nu_{\rm dyn}[\rm K_{\rm S}^{0},\rm K^{\pm}]$ is strongly violated. 

\begin{figure}
\centering
\includegraphics[width=8.0 cm,trim={5mm 2mm 4mm 1mm},clip] {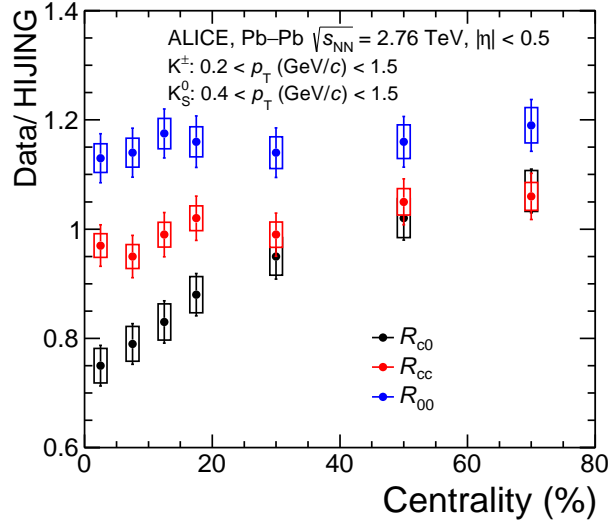}
\caption{Ratio (Data/HIJING) of individual terms of $\nu_{\rm dyn}[\rm K_{\rm S}^{0}, \rm K^{\pm}]$ as  a function of collision  centrality. Statistical and systematic uncertainties are represented as vertical  bar and boxes, respectively.} 
\label{fig:PID:TPC:kpkm}
\end{figure}

In order to interpret the dependence of $\nu_{\rm dyn}[\rm K_{\rm S}^{0},\rm K^{\pm}]$ on collision centrality and identify the origin of  the   $1/N_{\rm s}$  scaling violation,  we study the collision centrality dependence of the components $R_{\rm cc}$, $R_{\rm c0}$, and $R_{\rm 00}$ relative to those obtained with HIJING. Correlators computed with HIJING  have a nearly perfect  $1/N_{\rm s}$ scaling as a function of collision centrality. This means they can   be used as ``no-scaling-violation" baselines to  investigate the collision centrality dependence of  the evolution of the  measured $R_{\rm cc}$, $R_{\rm c0}$, and $R_{\rm 00}$ correlators with collision centrality. Figure~\ref{fig:PID:TPC:kpkm} presents  ratios of measured correlators to  those obtained with HIJING as a function of collision centrality.  
The ratio $R_{\rm 00}/R_{\rm 00}^{\rm HIJING}$ exhibits the largest deviation from unity but is otherwise independent, within uncertainties, of collision centrality. This term  thus essentially features the $1/N_{\rm s}$ scaling expected from a system consisting of a number of
independent sources, albeit with a magnitude larger by about 15\% than that expected from HIJING. By contrast, the  ratio $R_{\rm cc}/R_{\rm cc}^{\rm HIJING}$  is closest to unity but features a modest collision centrality dependence. This modest dependence, discussed further below, is not the main cause of the observed scaling violation of $\nu_{\rm dyn}[\rm K_{\rm S}^{0},\rm K^{\pm}]$ with collision centrality. Indeed, it is found that the ratio $R_{\rm c0}/R_{\rm c0}^{\rm HIJING}$ manifests a more significant collision centrality dependence. The ratio is of the order of unity in the  60--80\% collision centrality range with a deviation from unity consistent, more or less, with that observed for the ratio $R_{\rm cc}/R_{\rm cc}^{\rm HIJING}$  across all centralities. HIJING thus appears to provide a reasonable approximation of the measured correlation strength of neutral to charged kaons in peripheral collisions. However, the deviation of the measured $R_{\rm c0}$ from HIJING predictions increases monotonically from peripheral to central Pb--Pb collisions, with the largest deviation observed for the  0--5\% Pb--Pb collisions. We then conclude that it is the $R_{\rm c0}$ term  that most affect the collision centrality dependence and scaling violation of $\nu_{\rm dyn}[\rm K_{\rm S}^{0},\rm K^{\pm}]$.  Interestingly, it happens to be  the  term most sensitive to variations in the make up of kaons: combining a strange quark ($s$) with anti-up ($\bar u$) and anti-down quark  ($\bar d$), one obtains $\rm K^-$ and $\rm K^0$, respectively (similarly, combining $\bar s$ to $u$ and $d$ quarks, one obtains $\rm K^+$ and $\rm K^0$). Fluctuations in the relative number of neutral and charged kaons, measured by the $R_{\rm c0}$ term, are thus sensitive to fluctuations in the relative local abundances of $u$ ($\bar u$)  and $d$ ($\bar d$) quarks. The observed scaling violation  of $R_{\rm c0}$ with collision centrality thus suggests the relative abundances of $\bar u$ and $\bar d$ (as well as $u$ and $d$) available for the makeup of kaons might be evolving with collision centrality. 

The strength of charged kaon correlations is examined in closer detail  by plotting measured values of $\nu_{\rm dyn}[\rm K^{+},\rm K^{-}]$ and predictions 
by AMPT, HIJING and EPOS--LHC for this observable as a function of collision centrality in the bottom panel of Fig.~\ref{fig:NuDynVsCent} (a), and values scaled
by $\alpha$ in Fig.~\ref{fig:NuDynVsCent} (b). The deviations of measured $\nu_{\rm dyn}[\rm K^{+},\rm K^{-}]$ from model predictions are smaller than those for $\nu_{\rm dyn}[\rm K_{\rm S}^{0},\rm K^{\pm}]$ but 
measured $\nu_{\rm dyn}[\rm K^{+},\rm K^{-}]$ values lie systematically above those obtained from the models. 
 Although HIJING, AMPT and EPOS--LHC do not perfectly capture the magnitude and collision dependence of $\nu_{\rm dyn}[\rm K^{+},\rm K^{-}]$, they nonetheless provide a relatively accurate description of the role of charge  conservation, for charged kaons, 
in \PbPb\  collisions. Additionally note  that scaled values $\nu_{\rm dyn}[\rm K^{+},\rm K^{-}]/\alpha$ are invariant, within uncertainties,  with collision centrality, much like values obtained with the three models. Scaling violations of $\nu_{\rm dyn}[\rm K^{+},\rm K^{-}]$ with collision centrality, if any, are not observable within the uncertainties of this measurement   and stand in sharp contrast to the large scaling violation of  $\nu_{\rm dyn}[\rm K_{\rm S}^{0},\rm K^{\pm}]$ shown in the top panel of  Fig.~2(b). One can then conclude  that the large centrality dependence of $\nu_{\rm dyn}[\rm K_{\rm S}^{0},\rm K^{\pm}]$  does not 
originate from anomalous charge correlations.

The observed strong decrease of $R_{\rm c0}/R_{\rm c0}^{\rm HIJING}$ in central collisions indicates 
that the level of correlations between  neutral and charged kaons is weakening in this range relative to that predicted by HIJING. A weaker correlation is expected from the production of large strange DCCs, as shown in Fig. 5 of Ref.~\cite{PhysRevC.101.054904} presenting a simple phenomenological model of kaon DCC production. The observed dependence of $\nu_{\rm dyn}[\rm K_{\rm S}^{0},\rm K^{\pm}]$ on collision centrality is consistent with expectations based on a simple DCC model. However, significant contributions from other final state effects in central collisions might also dilute the correlation developed in initial stages.

 Nominally, the production of strange DCCs should manifest itself by the emission of relatively low-$p_{\rm T}$ kaons in the rest frame of the DCC~\cite{Gavin:2002ci}. 
 Even though there is no quantitative prediction in the literature, the radial acceleration~\cite{Kolb:2003dz,Retiere:2003kf} known to occur in relativistic A–A collisions may affect the DCCs and the particles they produce.  Therefore, the strength of the $\nu_{\rm dyn}[\rm K_{\rm S}^{0}, \rm K^{\pm}]$ correlator is studied in different ranges of transverse momentum. 

 \begin{figure}
\centering
\includegraphics[width=8.0 cm,trim={5mm 2mm 4mm 1mm},clip] {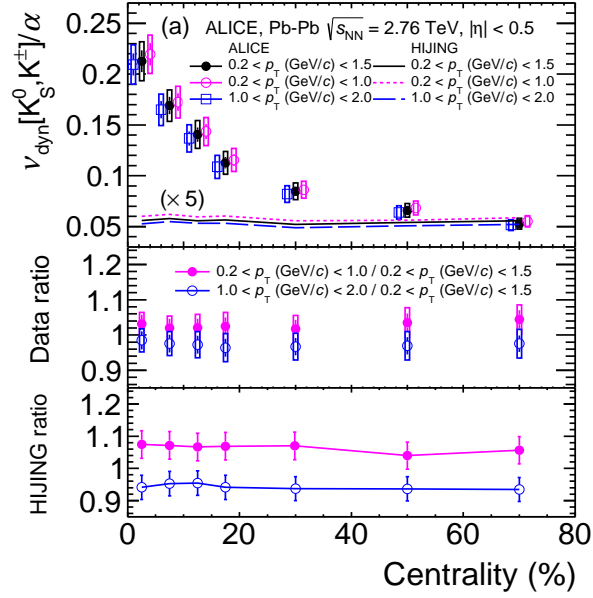}
\caption{(a) top: measured and HIJING (generator level) predicted values of $\nu_{\rm dyn}[\rm K_{\rm S}^{0},\rm K^{\pm}]$  scaled by $\alpha$ as a function of  collision centrality for various transverse momentum interval of charged kaons. HIJING predicted scaled values are multiplied by a factor of 5 to approximately match the measured values in most peripheral collisions; middle: ratio of data for various transverse momentum interval as a function of centrality; bottom: ratio of HIJING for various transverse momentum interval as a function of centrality.}
\label{fig:NuDynDifferentialpt}
\end{figure}

\begin{figure}
\centering
\includegraphics[width=0.49\linewidth,trim={3mm 2mm 4mm 1mm},clip] {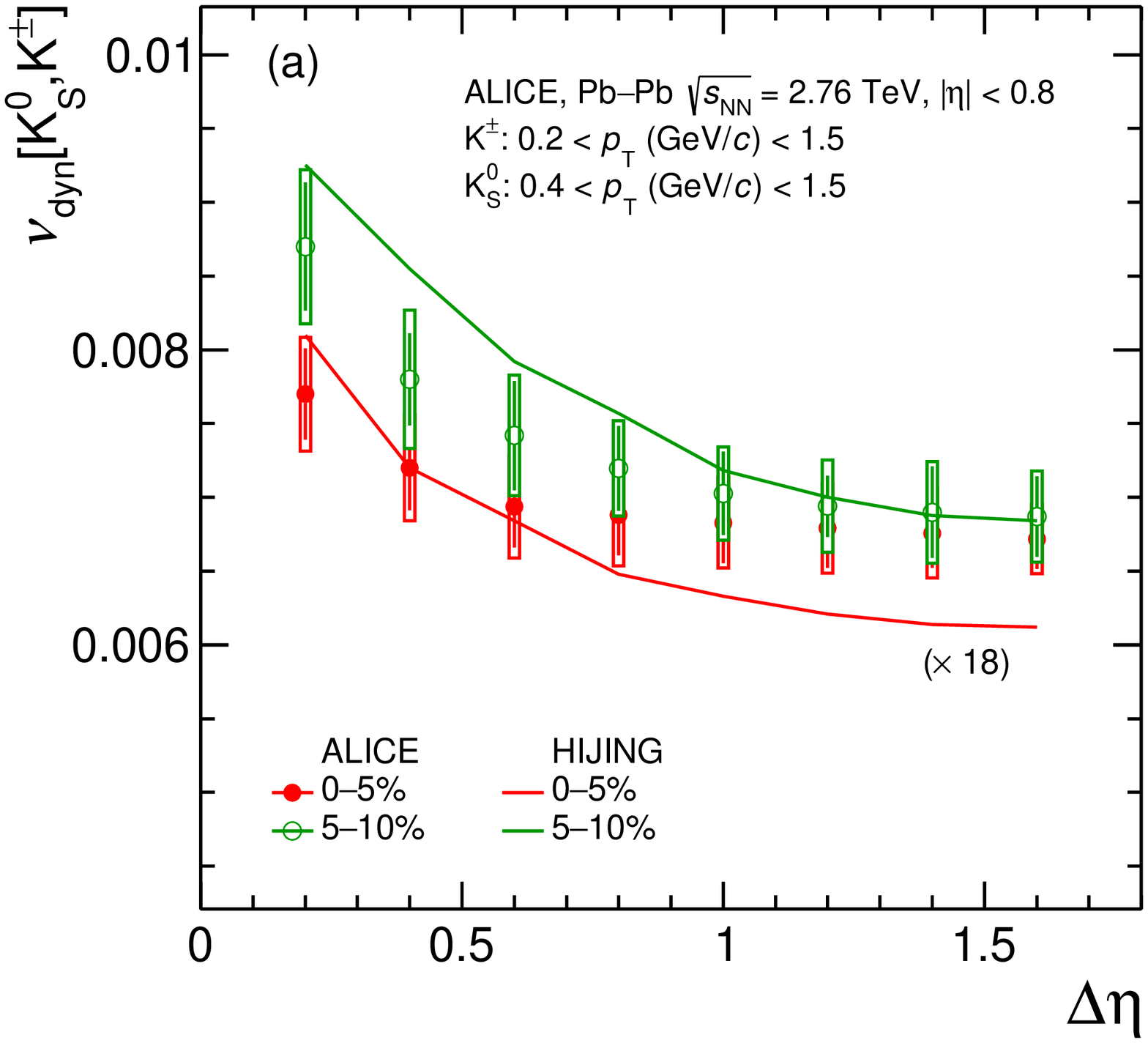}
\includegraphics[width=0.49\linewidth,trim={1mm 2mm 4mm 2mm},clip] {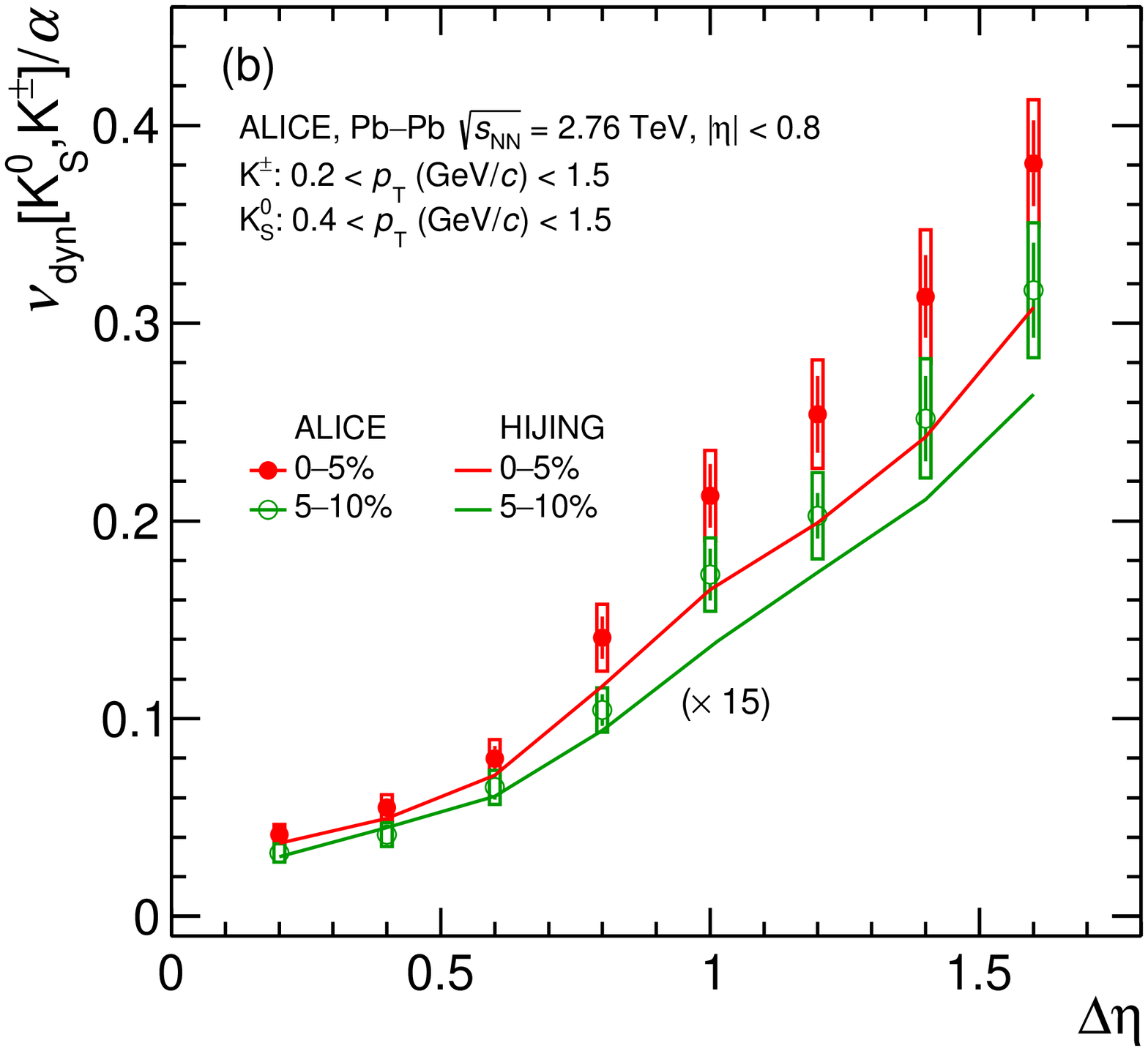}
\caption{(a) Measured values of $\nu_{\rm dyn}[\rm K_{\rm S}^{0},\rm K^{\pm}]$ plotted as a function of the width of the acceptance  $\Delta\eta$ in the 0--5\% and 5--10\% collision centrality ranges are compared  HIJING (generator level) calculations scaled by a factor of 18. (b) Values of $\nu_{\rm dyn}[\rm K_{\rm S}^{0},\rm K^{\pm}]$ shown in panel (a) are scaled by $\alpha$ . HIJING values are scaled by a factor of 15 for easier comparison with the data. The statistical and systematic uncertainties are represented as bar and boxes, respectively.}
\label{fig:NuDynDifferentialeta}
\end{figure}

  The collision centrality evolution of the scaled values of $\nu_{\rm dyn}[\rm K_{\rm S}^{0},\rm K^{\pm}]$ is shown in Fig.~\ref{fig:NuDynDifferentialpt} for selected $p_{\rm T}$ ranges. The $\nu_{\rm dyn}/\alpha$ scaling violation is observed in both the higher and lower selected $p_{\rm T}$ ranges. Contrary to what one expects from DCC production, within uncertainties, the scaled
correlation strength does not show a significant enhancement at low $p_{\rm T}$. As shown in the bottom panel of Fig.~\ref{fig:NuDynDifferentialpt}, HIJING also predicts larger correlation strengths in the lower $p_{\rm T}$ range. 
Additionally,    
Fig.~\ref{fig:NuDynDifferentialeta} 
presents the dependence of $\nu_{\rm dyn}[\rm K_{\rm S}^{0},\rm K^{\pm}]$, in panel (a), and  $\nu_{\rm dyn}[\rm K_{\rm S}^{0},\rm K^{\pm}]/\alpha$, in panel (b), on the width of the pseudorapidity acceptance, $\Delta \eta$, for 0--5\% and 5--10\% \PbPb\ collision centrality ranges. In panel (a), the data exhibit a monotonic decrease of $\nu_{\rm dyn}[\rm K_{\rm S}^{0},\rm K^{\pm}]$ with increasing $\Delta \eta$, which  reflects the finite correlation width of all three integral correlator terms $R_{xy}$, whereas in panel (b), scaled values of $\nu_{\rm dyn}[\rm K_{\rm S}^{0},\rm K^{\pm}]$ exhibit a monotonically decreasing trend with decreasing $\Delta \eta$ acceptance that stems largely from the decrease of the integrated yield of kaons with shrinking $\Delta \eta$ acceptance.  DCCs are expected to produce relatively low $p_{\rm T}$ particles~\cite{PhysRevD.46.246,2003NuPhA.715..657A} and should thus be characterized by relatively narrow correlation functions in momentum space. Radial acceleration associated with the collision system expansion shall further narrow  the correlation dependence on $\Delta\eta$. We note, however, that the peak widths, $\sigma_{\Delta\eta}$, observed in data are not significantly smaller than those obtained with the  HIJING calculations. Qualitatively, one would expect radial flow to further reduce the $\sigma_{\Delta\eta}$ difference among low-$p_{\rm T}$ kaons produced by the decay of strange DCCs.  Unfortunately, the absence of a prediction of the effect of radial flow on DCCs and given that the observed widths are only slightly smaller than those estimated with HIJING, the measured data thus do not   make a compelling case for the production of strange DCCs in \PbPb\ collisions at the TeV scale. 

In this letter, we presented  measurements of event-by-event fluctuations of the relative yield  of the neutral and  charged kaons  in \PbPb\  collisions at $\sqrt{s_{\rm NN}}$ = 2.76 TeV based on the $\nu_{\rm dyn}$ observable.  The centrality dependence of $\nu_{\rm dyn}$ is observed to violate the $1/N_{\rm s}$ multiplicity scaling expected from a system of $N$ independent sources, but this effect is not reproduced by HIJING, AMPT, and EPOS--LHC models.  Close examination of the three terms of $\nu_{\rm dyn}$
reveals that the strength of correlations among charged kaons features a collision centrality dependence close
to that expected with these three models.  The $R_{\rm c0}$ cross-term, however, is found to weaken
considerably, in most central collisions, relative to  a naive $1/N_{\rm s}$  expectation. This
indicates  correlations between charged and neutral 
kaons are significantly suppressed in central collisions. Given the fact that at this time it is unknown if other processes could mimic the signature of kaon production via DCCs and that the expected momentum dependence is not observed in the data, the reported measurement does not support the case for strange DCC production in heavy-ion collisions at LHC energies. Further measurements of  differential correlations in $\Delta \eta$ vs. $\Delta\phi$ and higher factorial moments are of interest, as they could provide information on  the momentum correlation length and the typical size of correlated kaon sources.

\newenvironment{acknowledgement}{\relax}{\relax}
\begin{acknowledgement}

The ALICE Collaboration would like to thank all its engineers and technicians for their invaluable contributions to the construction of the experiment and the CERN accelerator teams for the outstanding performance of the LHC complex.
The ALICE Collaboration gratefully acknowledges the resources and support provided by all Grid centres and the Worldwide LHC Computing Grid (WLCG) collaboration.
The ALICE Collaboration acknowledges the following funding agencies for their support in building and running the ALICE detector:
A. I. Alikhanyan National Science Laboratory (Yerevan Physics Institute) Foundation (ANSL), State Committee of Science and World Federation of Scientists (WFS), Armenia;
Austrian Academy of Sciences, Austrian Science Fund (FWF): [M 2467-N36] and Nationalstiftung f\"{u}r Forschung, Technologie und Entwicklung, Austria;
Ministry of Communications and High Technologies, National Nuclear Research Center, Azerbaijan;
Conselho Nacional de Desenvolvimento Cient\'{\i}fico e Tecnol\'{o}gico (CNPq), Financiadora de Estudos e Projetos (Finep), Funda\c{c}\~{a}o de Amparo \`{a} Pesquisa do Estado de S\~{a}o Paulo (FAPESP) and Universidade Federal do Rio Grande do Sul (UFRGS), Brazil;
Ministry of Education of China (MOEC) , Ministry of Science \& Technology of China (MSTC) and National Natural Science Foundation of China (NSFC), China;
Ministry of Science and Education and Croatian Science Foundation, Croatia;
Centro de Aplicaciones Tecnol\'{o}gicas y Desarrollo Nuclear (CEADEN), Cubaenerg\'{\i}a, Cuba;
Ministry of Education, Youth and Sports of the Czech Republic, Czech Republic;
The Danish Council for Independent Research | Natural Sciences, the VILLUM FONDEN and Danish National Research Foundation (DNRF), Denmark;
Helsinki Institute of Physics (HIP), Finland;
Commissariat \`{a} l'Energie Atomique (CEA) and Institut National de Physique Nucl\'{e}aire et de Physique des Particules (IN2P3) and Centre National de la Recherche Scientifique (CNRS), France;
Bundesministerium f\"{u}r Bildung und Forschung (BMBF) and GSI Helmholtzzentrum f\"{u}r Schwerionenforschung GmbH, Germany;
General Secretariat for Research and Technology, Ministry of Education, Research and Religions, Greece;
National Research, Development and Innovation Office, Hungary;
Department of Atomic Energy Government of India (DAE), Department of Science and Technology, Government of India (DST), University Grants Commission, Government of India (UGC) and Council of Scientific and Industrial Research (CSIR), India;
Indonesian Institute of Science, Indonesia;
Istituto Nazionale di Fisica Nucleare (INFN), Italy;
Japanese Ministry of Education, Culture, Sports, Science and Technology (MEXT) and Japan Society for the Promotion of Science (JSPS) KAKENHI, Japan;
Consejo Nacional de Ciencia (CONACYT) y Tecnolog\'{i}a, through Fondo de Cooperaci\'{o}n Internacional en Ciencia y Tecnolog\'{i}a (FONCICYT) and Direcci\'{o}n General de Asuntos del Personal Academico (DGAPA), Mexico;
Nederlandse Organisatie voor Wetenschappelijk Onderzoek (NWO), Netherlands;
The Research Council of Norway, Norway;
Commission on Science and Technology for Sustainable Development in the South (COMSATS), Pakistan;
Pontificia Universidad Cat\'{o}lica del Per\'{u}, Peru;
Ministry of Education and Science, National Science Centre and WUT ID-UB, Poland;
Korea Institute of Science and Technology Information and National Research Foundation of Korea (NRF), Republic of Korea;
Ministry of Education and Scientific Research, Institute of Atomic Physics, Ministry of Research and Innovation and Institute of Atomic Physics and University Politehnica of Bucharest, Romania;
Joint Institute for Nuclear Research (JINR), Ministry of Education and Science of the Russian Federation, National Research Centre Kurchatov Institute, Russian Science Foundation and Russian Foundation for Basic Research, Russia;
Ministry of Education, Science, Research and Sport of the Slovak Republic, Slovakia;
National Research Foundation of South Africa, South Africa;
Swedish Research Council (VR) and Knut \& Alice Wallenberg Foundation (KAW), Sweden;
European Organization for Nuclear Research, Switzerland;
Suranaree University of Technology (SUT), National Science and Technology Development Agency (NSDTA) and Office of the Higher Education Commission under NRU project of Thailand, Thailand;
Turkish Energy, Nuclear and Mineral Research Agency (TENMAK), Turkey;
National Academy of  Sciences of Ukraine, Ukraine;
Science and Technology Facilities Council (STFC), United Kingdom;
National Science Foundation of the United States of America (NSF) and United States Department of Energy, Office of Nuclear Physics (DOE NP), United States of America.    
\end{acknowledgement}
\bibliography{paper}
\bibliographystyle{utphys} 
\newpage
\appendix
\section{The ALICE Collaboration}
\label{app:collab}
\small
\begin{flushleft} 

S.~Acharya$^{\rm 142}$, 
D.~Adamov\'{a}$^{\rm 96}$, 
A.~Adler$^{\rm 74}$, 
J.~Adolfsson$^{\rm 81}$, 
G.~Aglieri Rinella$^{\rm 34}$, 
M.~Agnello$^{\rm 30}$, 
N.~Agrawal$^{\rm 54}$, 
Z.~Ahammed$^{\rm 142}$, 
S.~Ahmad$^{\rm 16}$, 
S.U.~Ahn$^{\rm 76}$
% ...

\section*{Affiliation Notes}

$^{\rm I}$ Deceased\\
$^{\rm II}$ Also at: Italian National Agency for New Technologies, Energy and Sustainable Economic Development (ENEA), Bologna, Italy\\
$^{\rm III}$ Also at: Dipartimento DET del Politecnico di Torino, Turin, Italy\\
$^{\rm IV}$ Also at: M.V. Lomonosov Moscow State University, D.V. Skobeltsyn Institute of Nuclear, Physics, Moscow, Russia\\
$^{\rm V}$ Also at: Department of Applied Physics, Aligarh Muslim University, Aligarh, India\\
$^{\rm VI}$ Also at: Institute of Theoretical Physics, University of Wroclaw, Poland\\
$^{\rm VII}$ Also at: University of Kansas, Lawrence, Kansas, United States\\

\section*{Collaboration Institutes}

$^{1}$ A.I. Alikhanyan National Science Laboratory (Yerevan Physics Institute) Foundation, Yerevan, Armenia\\
$^{2}$ AGH University of Science and Technology, Cracow, Poland\\
$^{3}$ Bogolyubov Institute for Theoretical Physics, National Academy of Sciences of Ukraine, Kiev, Ukraine\\
$^{4}$ Bose Institute, Department of Physics  and Centre for Astroparticle Physics and Space Science (CAPSS), Kolkata, India\\
$^{5}$ Budker Institute for Nuclear Physics, Novosibirsk, Russia\\
$^{6}$ California Polytechnic State University, San Luis Obispo, California, United States\\
$^{7}$ Central China Normal University, Wuhan, China\\
$^{8}$ Centro de Aplicaciones Tecnol\'{o}gicas y Desarrollo Nuclear (CEADEN), Havana, Cuba\\
$^{9}$ Centro de Investigaci\'{o}n y de Estudios Avanzados (CINVESTAV), Mexico City and M\'{e}rida, Mexico\\
$^{10}$ Chicago State University, Chicago, Illinois, United States\\
$^{11}$ China Institute of Atomic Energy, Beijing, China\\
$^{12}$ Chungbuk National University, Cheongju, Republic of Korea\\
$^{13}$ Comenius University Bratislava, Faculty of Mathematics, Physics and Informatics, Bratislava, Slovakia\\
$^{14}$ COMSATS University Islamabad, Islamabad, Pakistan\\
$^{15}$ Creighton University, Omaha, Nebraska, United States\\
$^{16}$ Department of Physics, Aligarh Muslim University, Aligarh, India\\
$^{17}$ Department of Physics, Pusan National University, Pusan, Republic of Korea\\
$^{18}$ Department of Physics, Sejong University, Seoul, Republic of Korea\\
$^{19}$ Department of Physics, University of California, Berkeley, California, United States\\
$^{20}$ Department of Physics, University of Oslo, Oslo, Norway\\
$^{21}$ Department of Physics and Technology, University of Bergen, Bergen, Norway\\
$^{22}$ Dipartimento di Fisica dell'Universit\`{a} and Sezione INFN, Cagliari, Italy\\
$^{23}$ Dipartimento di Fisica dell'Universit\`{a} and Sezione INFN, Trieste, Italy\\
$^{24}$ Dipartimento di Fisica dell'Universit\`{a} and Sezione INFN, Turin, Italy\\
$^{25}$ Dipartimento di Fisica e Astronomia dell'Universit\`{a} and Sezione INFN, Bologna, Italy\\
$^{26}$ Dipartimento di Fisica e Astronomia dell'Universit\`{a} and Sezione INFN, Catania, Italy\\
$^{27}$ Dipartimento di Fisica e Astronomia dell'Universit\`{a} and Sezione INFN, Padova, Italy\\
$^{28}$ Dipartimento di Fisica e Nucleare e Teorica, Universit\`{a} di Pavia, Pavia, Italy\\
$^{29}$ Dipartimento di Fisica `E.R.~Caianiello' dell'Universit\`{a} and Gruppo Collegato INFN, Salerno, Italy\\
$^{30}$ Dipartimento DISAT del Politecnico and Sezione INFN, Turin, Italy\\
$^{31}$ Dipartimento di Scienze e Innovazione Tecnologica dell'Universit\`{a} del Piemonte Orientale and INFN Sezione di Torino, Alessandria, Italy\\
$^{32}$ Dipartimento di Scienze MIFT, Universit\`{a} di Messina, Messina, Italy\\
$^{33}$ Dipartimento Interateneo di Fisica `M.~Merlin' and Sezione INFN, Bari, Italy\\
$^{34}$ European Organization for Nuclear Research (CERN), Geneva, Switzerland\\
$^{35}$ Faculty of Electrical Engineering, Mechanical Engineering and Naval Architecture, University of Split, Split, Croatia\\
$^{36}$ Faculty of Engineering and Science, Western Norway University of Applied Sciences, Bergen, Norway\\
$^{37}$ Faculty of Nuclear Sciences and Physical Engineering, Czech Technical University in Prague, Prague, Czech Republic\\
$^{38}$ Faculty of Science, P.J.~\v{S}af\'{a}rik University, Ko\v{s}ice, Slovakia\\
$^{39}$ Frankfurt Institute for Advanced Studies, Johann Wolfgang Goethe-Universit\"{a}t Frankfurt, Frankfurt, Germany\\
$^{40}$ Fudan University, Shanghai, China\\
$^{41}$ Gangneung-Wonju National University, Gangneung, Republic of Korea\\
$^{42}$ Gauhati University, Department of Physics, Guwahati, India\\
$^{43}$ Helmholtz-Institut f\"{u}r Strahlen- und Kernphysik, Rheinische Friedrich-Wilhelms-Universit\"{a}t Bonn, Bonn, Germany\\
$^{44}$ Helsinki Institute of Physics (HIP), Helsinki, Finland\\
$^{45}$ High Energy Physics Group,  Universidad Aut\'{o}noma de Puebla, Puebla, Mexico\\
$^{46}$ Hiroshima University, Hiroshima, Japan\\
$^{47}$ Hochschule Worms, Zentrum  f\"{u}r Technologietransfer und Telekommunikation (ZTT), Worms, Germany\\
$^{48}$ Horia Hulubei National Institute of Physics and Nuclear Engineering, Bucharest, Romania\\
$^{49}$ Indian Institute of Technology Bombay (IIT), Mumbai, India\\
$^{50}$ Indian Institute of Technology Indore, Indore, India\\
$^{51}$ Indonesian Institute of Sciences, Jakarta, Indonesia\\
$^{52}$ INFN, Laboratori Nazionali di Frascati, Frascati, Italy\\
$^{53}$ INFN, Sezione di Bari, Bari, Italy\\
$^{54}$ INFN, Sezione di Bologna, Bologna, Italy\\
$^{55}$ INFN, Sezione di Cagliari, Cagliari, Italy\\
$^{56}$ INFN, Sezione di Catania, Catania, Italy\\
$^{57}$ INFN, Sezione di Padova, Padova, Italy\\
$^{58}$ INFN, Sezione di Pavia, Pavia, Italy\\
$^{59}$ INFN, Sezione di Torino, Turin, Italy\\
$^{60}$ INFN, Sezione di Trieste, Trieste, Italy\\
$^{61}$ Inha University, Incheon, Republic of Korea\\
$^{62}$ Institute for Gravitational and Subatomic Physics (GRASP), Utrecht University/Nikhef, Utrecht, Netherlands\\
$^{63}$ Institute for Nuclear Research, Academy of Sciences, Moscow, Russia\\
$^{64}$ Institute of Experimental Physics, Slovak Academy of Sciences, Ko\v{s}ice, Slovakia\\
$^{65}$ Institute of Physics, Homi Bhabha National Institute, Bhubaneswar, India\\
$^{66}$ Institute of Physics of the Czech Academy of Sciences, Prague, Czech Republic\\
$^{67}$ Institute of Space Science (ISS), Bucharest, Romania\\
$^{68}$ Institut f\"{u}r Kernphysik, Johann Wolfgang Goethe-Universit\"{a}t Frankfurt, Frankfurt, Germany\\
$^{69}$ Instituto de Ciencias Nucleares, Universidad Nacional Aut\'{o}noma de M\'{e}xico, Mexico City, Mexico\\
$^{70}$ Instituto de F\'{i}sica, Universidade Federal do Rio Grande do Sul (UFRGS), Porto Alegre, Brazil\\
$^{71}$ Instituto de F\'{\i}sica, Universidad Nacional Aut\'{o}noma de M\'{e}xico, Mexico City, Mexico\\
$^{72}$ iThemba LABS, National Research Foundation, Somerset West, South Africa\\
$^{73}$ Jeonbuk National University, Jeonju, Republic of Korea\\
$^{74}$ Johann-Wolfgang-Goethe Universit\"{a}t Frankfurt Institut f\"{u}r Informatik, Fachbereich Informatik und Mathematik, Frankfurt, Germany\\
$^{75}$ Joint Institute for Nuclear Research (JINR), Dubna, Russia\\
$^{76}$ Korea Institute of Science and Technology Information, Daejeon, Republic of Korea\\
$^{77}$ KTO Karatay University, Konya, Turkey\\
$^{78}$ Laboratoire de Physique des 2 Infinis, Ir\`{e}ne Joliot-Curie, Orsay, France\\
$^{79}$ Laboratoire de Physique Subatomique et de Cosmologie, Universit\'{e} Grenoble-Alpes, CNRS-IN2P3, Grenoble, France\\
$^{80}$ Lawrence Berkeley National Laboratory, Berkeley, California, United States\\
$^{81}$ Lund University Department of Physics, Division of Particle Physics, Lund, Sweden\\
$^{82}$ Moscow Institute for Physics and Technology, Moscow, Russia\\
$^{83}$ Nagasaki Institute of Applied Science, Nagasaki, Japan\\
$^{84}$ Nara Women{'}s University (NWU), Nara, Japan\\
$^{85}$ National and Kapodistrian University of Athens, School of Science, Department of Physics , Athens, Greece\\
$^{86}$ National Centre for Nuclear Research, Warsaw, Poland\\
$^{87}$ National Institute of Science Education and Research, Homi Bhabha National Institute, Jatni, India\\
$^{88}$ National Nuclear Research Center, Baku, Azerbaijan\\
$^{89}$ National Research Centre Kurchatov Institute, Moscow, Russia\\
$^{90}$ Niels Bohr Institute, University of Copenhagen, Copenhagen, Denmark\\
$^{91}$ Nikhef, National institute for subatomic physics, Amsterdam, Netherlands\\
$^{92}$ NRC Kurchatov Institute IHEP, Protvino, Russia\\
$^{93}$ NRC \guillemotleft Kurchatov\guillemotright  Institute - ITEP, Moscow, Russia\\
$^{94}$ NRNU Moscow Engineering Physics Institute, Moscow, Russia\\
$^{95}$ Nuclear Physics Group, STFC Daresbury Laboratory, Daresbury, United Kingdom\\
$^{96}$ Nuclear Physics Institute of the Czech Academy of Sciences, \v{R}e\v{z} u Prahy, Czech Republic\\
$^{97}$ Oak Ridge National Laboratory, Oak Ridge, Tennessee, United States\\
$^{98}$ Ohio State University, Columbus, Ohio, United States\\
$^{99}$ Petersburg Nuclear Physics Institute, Gatchina, Russia\\
$^{100}$ Physics department, Faculty of science, University of Zagreb, Zagreb, Croatia\\
$^{101}$ Physics Department, Panjab University, Chandigarh, India\\
$^{102}$ Physics Department, University of Jammu, Jammu, India\\
$^{103}$ Physics Department, University of Rajasthan, Jaipur, India\\
$^{104}$ Physikalisches Institut, Eberhard-Karls-Universit\"{a}t T\"{u}bingen, T\"{u}bingen, Germany\\
$^{105}$ Physikalisches Institut, Ruprecht-Karls-Universit\"{a}t Heidelberg, Heidelberg, Germany\\
$^{106}$ Physik Department, Technische Universit\"{a}t M\"{u}nchen, Munich, Germany\\
$^{107}$ Politecnico di Bari and Sezione INFN, Bari, Italy\\
$^{108}$ Research Division and ExtreMe Matter Institute EMMI, GSI Helmholtzzentrum f\"ur Schwerionenforschung GmbH, Darmstadt, Germany\\
$^{109}$ Russian Federal Nuclear Center (VNIIEF), Sarov, Russia\\
$^{110}$ Saha Institute of Nuclear Physics, Homi Bhabha National Institute, Kolkata, India\\
$^{111}$ School of Physics and Astronomy, University of Birmingham, Birmingham, United Kingdom\\
$^{112}$ Secci\'{o}n F\'{\i}sica, Departamento de Ciencias, Pontificia Universidad Cat\'{o}lica del Per\'{u}, Lima, Peru\\
$^{113}$ St. Petersburg State University, St. Petersburg, Russia\\
$^{114}$ Stefan Meyer Institut f\"{u}r Subatomare Physik (SMI), Vienna, Austria\\
$^{115}$ SUBATECH, IMT Atlantique, Universit\'{e} de Nantes, CNRS-IN2P3, Nantes, France\\
$^{116}$ Suranaree University of Technology, Nakhon Ratchasima, Thailand\\
$^{117}$ Technical University of Ko\v{s}ice, Ko\v{s}ice, Slovakia\\
$^{118}$ The Henryk Niewodniczanski Institute of Nuclear Physics, Polish Academy of Sciences, Cracow, Poland\\
$^{119}$ The University of Texas at Austin, Austin, Texas, United States\\
$^{120}$ Universidad Aut\'{o}noma de Sinaloa, Culiac\'{a}n, Mexico\\
$^{121}$ Universidade de S\~{a}o Paulo (USP), S\~{a}o Paulo, Brazil\\
$^{122}$ Universidade Estadual de Campinas (UNICAMP), Campinas, Brazil\\
$^{123}$ Universidade Federal do ABC, Santo Andre, Brazil\\
$^{124}$ University of Cape Town, Cape Town, South Africa\\
$^{125}$ University of Houston, Houston, Texas, United States\\
$^{126}$ University of Jyv\"{a}skyl\"{a}, Jyv\"{a}skyl\"{a}, Finland\\
$^{127}$ University of Kansas, Lawrence, Kansas, United States\\
$^{128}$ University of Liverpool, Liverpool, United Kingdom\\
$^{129}$ University of Science and Technology of China, Hefei, China\\
$^{130}$ University of South-Eastern Norway, Tonsberg, Norway\\
$^{131}$ University of Tennessee, Knoxville, Tennessee, United States\\
$^{132}$ University of the Witwatersrand, Johannesburg, South Africa\\
$^{133}$ University of Tokyo, Tokyo, Japan\\
$^{134}$ University of Tsukuba, Tsukuba, Japan\\
$^{135}$ University Politehnica of Bucharest, Bucharest, Romania\\
$^{136}$ Universit\'{e} Clermont Auvergne, CNRS/IN2P3, LPC, Clermont-Ferrand, France\\
$^{137}$ Universit\'{e} de Lyon, CNRS/IN2P3, Institut de Physique des 2 Infinis de Lyon, Lyon, France\\
$^{138}$ Universit\'{e} de Strasbourg, CNRS, IPHC UMR 7178, F-67000 Strasbourg, France, Strasbourg, France\\
$^{139}$ Universit\'{e} Paris-Saclay Centre d'Etudes de Saclay (CEA), IRFU, D\'{e}partment de Physique Nucl\'{e}aire (DPhN), Saclay, France\\
$^{140}$ Universit\`{a} degli Studi di Foggia, Foggia, Italy\\
$^{141}$ Universit\`{a} di Brescia, Brescia, Italy\\
$^{142}$ Variable Energy Cyclotron Centre, Homi Bhabha National Institute, Kolkata, India\\
$^{143}$ Warsaw University of Technology, Warsaw, Poland\\
$^{144}$ Wayne State University, Detroit, Michigan, United States\\
$^{145}$ Westf\"{a}lische Wilhelms-Universit\"{a}t M\"{u}nster, Institut f\"{u}r Kernphysik, M\"{u}nster, Germany\\
$^{146}$ Wigner Research Centre for Physics, Budapest, Hungary\\
$^{147}$ Yale University, New Haven, Connecticut, United States\\
$^{148}$ Yonsei University, Seoul, Republic of Korea\\

\end{flushleft} 
\end{document}